\documentstyle[multicol,aps,graphicx]{revtex}
\begin{document}
\draft
\title{State Orthogonalization by Building a Hilbert Space: A
New Approach to Electronic Quantum Transport in Molecular 
Wires\footnote{Copyright 1998 American Physical Society}}
 \author{Eldon Emberly and George 
Kirczenow} 
\address{Department of Physics, Simon Fraser University,
Burnaby, B.C., Canada  V5A 1S6}
\date{\today}
\maketitle
\begin{abstract}

Quantum descriptions of many complex systems
are formulated most naturally in bases of states
that are not mutually orthogonal. We introduce a
general and powerful yet simple approach that
facilitates solving such models exactly by embedding
the non-orthogonal states in a new Hilbert space in
which they are by definition mutually orthogonal. This
novel approach is applied to electronic transport
in molecular quantum wires and is used to predict
conductance antiresonances of a new type that arise
solely out of the non-orthogonality of the local orbitals
on different sites of the wire.

\end{abstract}
\begin{multicols}{2}

The predictions of quantum mechanics that relate to observable
phenomena do not depend on the particular basis that is
selected to represent state vectors in Hilbert space.  However
choosing a basis of states whose physical significance is clear
and that are mutually orthogonal can be extremely helpful in
theoretical work.  Bases constructed from the eigenstates of a
set of commuting operators that represent physical observables
have both of these desirable properties\cite{Dirac}. But for
complex systems that are composed of simpler building blocks,
it is tempting to use the eigenstates of the Hamiltonians of
the separate building blocks as basis states even though such
bases are not orthogonal.  In solid state physics and quantum
chemistry where the building blocks are atoms, each with its
own electronic eigenstates, this choice of basis gives rise to
the widely used tight-binding\cite{Ziman} and
H{\"u}ckel\cite{Hoff63} models.  An analogous approach that
models nucleons as bags of quarks\cite{nuclear} is used in
theoretical work on nuclei and nuclear matter.\cite{review} The
representations of the Hilbert spaces of complex systems that
are obtained in this way are intuitively appealing but the
non-orthogonality has been a significant drawback.  Standard
orthogonalization schemes such as Gram-Schmidt do not help here
because they are unwieldy for large systems and do not preserve
the atomistic character of the basis states.  Wannier
functions\cite{Ziman} provide orthogonalized local basis
states for perfect periodic structures, but they have the
disadvantage of not being eigenstates of atomic Hamiltonians and
also have no analog for disordered solids, liquids or molecules.
L\"{o}wdin functions\cite{Lowd50} do not require a periodic
lattice but they too are not eigenstates of atomic Hamiltonians
and do not have a simple physical interpretation. Thus, rather
than working with nonorthogonal bases,\cite{Fulde} it has been
customary in much of the literature, for the sake of simplicity,
to neglect the overlaps between the non-orthogonal states, as is
done in LCAO (linear combination of atomic orbitals) models of
electronic structure, tight-binding theories of Anderson
localization in disordered systems, and Hubbard and t-J models of
electronic correlations.  However, non-orthogonality can have
non-trivial physical implications. For example, it is closely
related to gauge interactions and fractional exclusion
statistics, as has been pointed out by Haldane\cite{Haldane}.
Better ways to treat it are therefore of interest.

In this Letter we introduce a simple, powerful and general
method that facilitates obtaining {\em exact} solutions of
quantum problems in which the non-orthogonality of the
basis is important. Our approach is {\em to build a new
Hilbert space} around the non-orthogonal basis states, a
space in which these states are {\em by definition}
orthogonal, and to work in this new Hilbert space. We
demonstrate the power, simplicity and flexibility of this
novel approach by applying it to electronic quantum
transport in molecular wires and predicting that these
should exhibit conductance antiresonances of a new type.
These antiresonances are a direct and surprising physical
consequence of the non-orthogonality of the electronic
orbitals on different atoms of the wire, which is
included naturally in our theory.

A molecular wire is a single molecule connecting a pair of
metallic contacts. Such devices have recently begun to be
realized experimentally and their electrical conductances are
being measured.\cite{Reed97,Datta97,Andres96,Bumm96} Electron
transport in molecular wires has been studied theoretically by
considering the transmission probability for electrons to
scatter through the
structure.\cite{Datta97,Samant96,Kemp96,Joach96,Mujic97,Ember98}
As with other mesoscopic systems, the electrical conductance
$G$ of the molecule is related to the transmission probability
$T$ at the Fermi level by the Landauer formula $G = {\frac
{e^2} h} T$.\cite{Lan57} Molecular wires display a number of
interesting transport phenomena, one of which is transmission
antiresonances.\cite{Ratner90,Cast93,Cheong94,Kemp96}
Antiresonances are zeroes of the transmission $T$ and
correspond to electrons being perfectly reflected by the
molecule. They also occur in semiconductor
systems.\cite{semicond} An analytic theory of molecular
antiresonances was first proposed by Ratner\cite{Ratner90} in
the context of electron transfer between donor and acceptor
sites of a molecule, but his treatment did not
include the effects of the non-orthogonality of the atomic
states. The present approach yields an analytic description of
molecular wire transport (including antiresonances) that treats
this non-orthogonality exactly and permits us to explore its
physical implications.

Our starting point is the Schr{\" o}dinger equation
\begin{equation}
H |\Psi\rangle  =  E |\Psi\rangle
\label{eq:Schrod}
\end{equation}
where the Hamiltonian operator $H$ and eigenvector
$|\Psi\rangle$ are defined in a Hilbert space $A$. We wish to
solve Eq.(\ref{eq:Schrod}) for $|\Psi\rangle$ which is
expressed as $|\Psi\rangle = \sum_n \Psi_n |n\rangle$ in a
non-orthogonal basis $\lbrace |n\rangle \rbrace$ of Hilbert
space $A$. In this basis Eq.(\ref{eq:Schrod}) takes the form
\begin{equation}
\sum_n H_{m,n} \Psi_n  =  E \sum_n S_{m,n} \Psi_n
\label{eq:Schrodbasis}
\end{equation}
where $H_{m,n}=\langle m |H|n\rangle$ and
$S_{m,n}=\langle m |n\rangle$ is the overlap
matrix.\cite{var}
Let us rewrite Eq.(\ref{eq:Schrodbasis}) as
\begin{equation}
\sum_n H^E_{m,n} \Psi_n  =  E \Psi_m
\label{eq:Schrodbasis'}
\end{equation}
where
\begin{equation}
H^E_{m,n} = H_{m,n} - E(S_{m,n}-\delta_{m,n}).
\label{eq:H'}
\end{equation}
Eq.(\ref{eq:Schrodbasis'}) can be viewed as
the matrix form of the Schr{\" o}dinger equation
\begin{equation}
H^E |\Psi'\rangle  =  E |\Psi'\rangle
\label{eq:Schrod'}
\end{equation}
where $H^E$ and $|\Psi'\rangle$ are a new Hamiltonian operator
and its eigenvector defined in a new Hilbert space $A'$
\cite{hilbert} in which the basis states $\lbrace |n\rangle \rbrace$
are {\em orthonormal} with $\langle m |n\rangle=\delta_{m,n}
\,$,
$|\Psi'\rangle = \sum_n \Psi_n |n\rangle$ has the {\em same}
coefficients $\Psi_n$ as $|\Psi\rangle$, and the new
Hamiltonian operator is defined by its matrix elements through
$\langle m |H^E|n\rangle = H^E_{m,n}$.  According to
Eq.(\ref{eq:H'}), $H^E$ is Hermitian in $A'$ because $E$ is
real and $H_{m,n}\,$, $S_{m,n}$ and $\delta_{m,n}$ are
Hermitian matrices.

Thus we have transformed a problem that was formulated in terms
of a nonorthogonal basis into an equivalent one in an
orthogonal basis in a {\em different} Hilbert space.
Other orthogonalization schemes (such as that of L\"{o}wdin
\cite{Lowd50}) differ from ours in this as well as other
respects.\cite{difference}

It
should be noted that only the eigenvectors of $H^E$ that have the
eigenvalue $E$ have the same coefficients $\Psi_n$ as
eigenvectors of the true Hamiltonian $H$. The other
eigenvectors of $H^E$ do not correspond to any eigenstate of
the physical Hamiltonian $H$, but they never the less play an
important role when calculating the Green's function
corresponding to $H^E$.

Since no assumptions at all have been made about the nature of
the system being considered, this method of orthogonalization
by switching to a new Hilbert space is extremely general. If
the basis states $\lbrace |n\rangle \rbrace$ are tight binding
atomic orbitals, then the present transformation (unlike the
transformation to Wannier functions) can be used irrespective
of the types of atoms involved or their locations in
space. Furthermore, our transformation has the additional
flexibility that the non-orthogonal basis states need not all
be of the same generic type. For example, some of them may be
atomic orbitals and others molecular orbitals on some
cluster(s) of atoms that form a part of the physical
system. This flexibility will be exploited below.  We now
proceed to outline the application to molecular wire quantum
transport.

We begin by solving analytically an idealized model of a
molecular wire consisting of a molecule attached to two
identical semi-infinite single-channel leads which are
represented by 1D chains of atoms. For this system we find it
convenient to choose a non-orthogonal basis consisting of
atomic orbitals $\{ |n\rangle \}$ with $n=-\infty \dots -1$ on
the left lead and $n=1 \dots \infty$ on the right lead and
molecular orbitals (MO's) $\{|j\rangle\}$ for the molecule.  In
terms of this basis we write electron eigenstates
$|\Psi\rangle$ of $H$ in which the electron is incident on the
molecule from the left lead and transmitted with probability
$T$ through the molecule to the right lead as
\begin{eqnarray}
|\Psi\rangle &=& \sum_{n=-\infty}^{-1} \Psi_{n} |n\rangle +
\sum_{n=1}^{\infty} \Psi_{n} |n\rangle +
\sum_j c_j |j\rangle
\label{eq:psi}
\end{eqnarray}
The transmission probability $T$ that enters the Landauer
electrical conductance of the wire is given by
$T=|\Psi_{1}|^2$. Solving for $\Psi_1$ analytically in the
nonorthogonal basis is difficult so we transform to the new
Hilbert space $A'$ where the solution is more straightforward.
As we have already shown, the coefficients $\Psi_{n}$ remain
the same in $A'$ where the basis
$\lbrace |n\rangle , |j\rangle \rbrace$ is orthogonal, so that
the above expression for $T$ is valid in either
Hilbert space.

We evaluate $\Psi_{1}$ by solving a Lippmann-Schwinger (LS)
equation that describes electron scattering in the molecular
wire.  We {\em define} this LS equation in the {\em new}
Hilbert space $A'$ where it takes the form
\begin{equation} |\Psi ' \rangle  =  |\Phi ' \rangle + G'(E)
W^E |\Psi ' \rangle .
\label{eq:LS}
\end{equation}
Here $|\Psi'\rangle$ is the eigenstate of the transformed
Hamiltonian $H^E$ whose coefficients correspond to those of the
untransformed scattering eigenstate $|\Psi\rangle$ defined
above Eq.(\ref {eq:psi}).
$W^E$ is defined by separating $H^E$ into two parts,
$H^E=H_o^E + W^E$, where the matrix elements of $H_o^E$
between lead orbitals
$\{|n\rangle\}$ and MO's $\{|j\rangle\}$ all
vanish {\em in the space}
$A'$ and
$W^E$ couples the molecule to the adjacent lead sites.  $|\Phi'
\rangle$ is an eigenstate of $H_0^E$ with eigenvalue $E$ that
is confined to the left lead.  $G'(E)=(E - H_0^E)^{-1}$ is the
Green's function of the decoupled system.

The validity of the LS Eq.(\ref {eq:LS}) depends
crucially on the clear distinction
between states on the leads and those on the molecule
that can only be made in Hilbert space $A'$;
non-orthogonality leads to contradictions if
analogs of the entities that enter (\ref {eq:LS}) are
constructed in $A$. The transformation to $A'$ also introduces
energy dependent hopping into the transformed Hamiltonian as
prescribed in Eq.  (\ref{eq:H'}). The energy dependence of the
coupling $W^E$ between the molecule and leads in Hilbert
space $A'$ will be important in the determination of
antiresonances. Our choice of a set of MO's
$\{|j\rangle\}$ that are mutually orthogonal
in $A$ means that the Green's function
for the isolated molecule is formally unaffected by our
transformation. This choice allows a simpler evaluation
of the molecular Green's function that enters $G'(E)$.  An
atomic orbital basis set non-orthogonal in $A$ could be chosen
instead, however the solution would then be less transparent.

The evaluation of the Green's function $G'(E)$ and solution of
the LS equation (\ref{eq:LS}) in the orthogonal basis $\lbrace
|n\rangle , |j\rangle \rbrace$ of Hilbert space $A'$ is
straightforward and will be presented in detail
elsewhere. Here we will focus on the salient results for
molecular wire conductance antiresonances and their physical
significance. We find
\begin{equation}
\Psi_{1} = \frac{P \Phi'_{-1}}{[(1-Q)(1-R)-PS]} \label{eq:psi1}
\end{equation}
where
\begin{eqnarray*}
P &=& G'_{1,1} \sum_j W^E_{1,j} G'_j W^E_{j,-1} \\
Q &=& G'_{1,1} \sum_j (W^E_{1,j})^2 G'_j \\
R &=& G'_{1,1} \sum_j (W^E_{-1,j})^2 G'_j \\
S &=& G'_{1,1} \sum_j W^E_{-1,j} G'_j W^E_{j,1}
\end{eqnarray*}
The sum over $j$ is over only the MO's.  In the above,
$W^E_{1,j} = H_{1,j} - E S_{1,j}$ is the energy-dependent
hopping element of $H^E$ between the first lead site and the
$j^{th}$ MO in terms of the hopping element of the original
Hamiltonian $H$ and the overlap in the non-orthogonal
basis. The Green's function on the molecule is expanded in
terms of its molecular eigenstates (which, as mentioned above,
are unchanged by the transformation) and this gives $G'_j =
1/(E-\epsilon_j)$ for the $j^{th}$ MO with energy
$\epsilon_j$. $G'_{1,1}$ is the diagonal matrix element of the
Green's function $G'(E)$ at the end site of the isolated lead.

Conductance antiresonances of the molecular wire occur
where the transmission $T=|\Psi_1|^2 $ is equal to zero. From
Eq. (\ref{eq:psi1}) this happens when
$P=0$, i.e., at Fermi energies E that are the roots of
\begin{equation}
\sum_j \frac{(H_{1,j} - E S_{1,j})(H_{j,-1} - E S_{j,-1})}{E -
\epsilon_j} = 0.
\label{eq:anti}
\end{equation}

Two distinct mechanisms for antiresonances in molecular wire
transport can be identified from Eq. (\ref{eq:anti}):

In the first of these mechanisms, antiresonances arise due to
an interference between molecular states that may differ in
energy, as is seen directly from Eq. (\ref{eq:anti}): An
electron incident from the left lead, hops from the lead site
adjacent to the molecule onto each of the MO's with a weight
$W^E_{j,-1}$. It propagates through each
of the different orbitals $j$ and hops onto the right lead with
a weight $W^E_{1,j} \,.$ These processes interfere with each
other and where they cancel (\ref{eq:anti}) is satisfied and an
antiresonance occurs.  This is in essence the same interference
mechanism as has been identified previously in work on electron
transfer between molecular donor and acceptor sites and, in the
absence of the overlaps $S_{1,j}$ and $S_{j,-1}$, our result
(\ref{eq:anti}) agrees with that obtained there.\cite{Ratner90}

The second antiresonance mechanism, which has no analog in
previous work, arises solely from the nonorthogonality of
atomic orbitals that we have included in our analytic theory
with the help of the Hilbert space transformation. It occurs
when only a single MO $k$ couples appreciably to the leads.  In
such cases, Eq.  (\ref{eq:anti}) becomes $(H_{k,-1} - E
S_{k,-1})(H_{1,k} - E S_{1,k}) = 0$.  Two antiresonances are
possible in this case. They occur at energies $E$ where a
matrix element of the transformed Hamiltonian ${\bf H}^E = {\bf
H} - E({\bf S-1})$ that is responsible for hopping between the
molecule and one of the leads vanishes in the new Hilbert space.
Thus the non-orthogonality of two orbitals can actually {\em
prevent} electron hopping between these orbitals from taking
place, blocking electron transmission along the wire and creating
an antiresonance. This is a counterintuitive effect since one
would normally expect orbital overlap to aid electron transfer
between the orbitals rather than hinder it. It underscores the
importance of including the effects of nonorthogonality fully in
tight-binding theories.

\begin{figure}[ht]
\includegraphics[bb = 0 0 640 800, width = 
0.75\textwidth,clip]{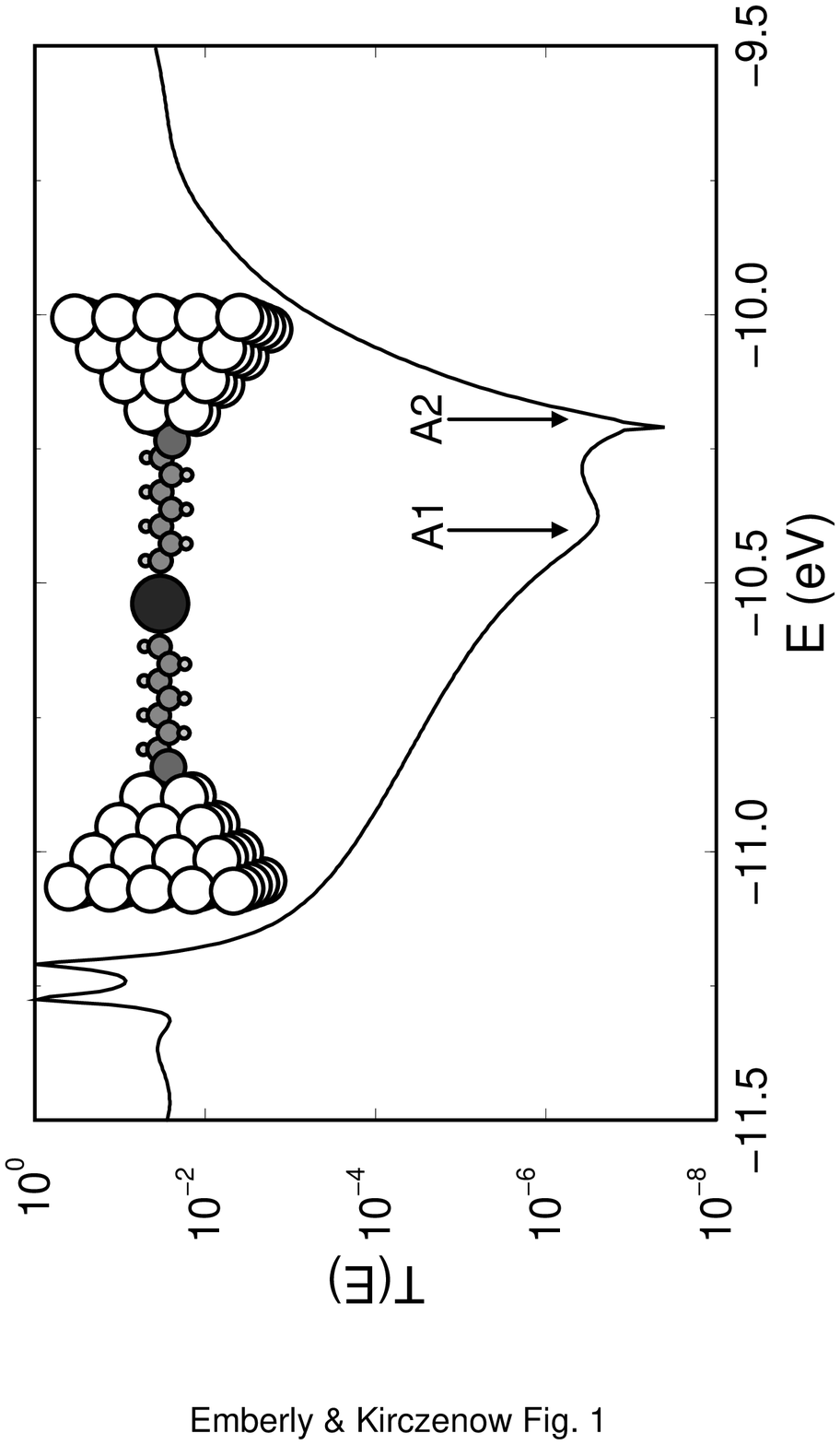}
\begin{center}
\caption{ Electronic transmission probability $T$ calculated
numerically for molecular wire shown in inset.  Antiresonance
is predicted at A1 and A2 by the analytic condition
(\ref{eq:anti}). The active molecule has two
$\pi$ levels with energies
$\epsilon_a = -13.0$ eV and $\epsilon_b = -9.0$ eV. The
coupling and overlap of these two levels to the left (L)
and right (R) chain molecules are
$W_{L,a}=-5.0$ eV, $W_{L,b}=-2.5$ eV,
$S_{L,a}=0.3$, $S_{L,b}=0.2$ and $W_{R,a}=-2.7$ eV,
$W_{R,b}=-1.8$ eV, $S_{R,a}=0.25$, $S_{R,b} = 0.15$.
}
\end{center}
\label{fig1}
\end{figure}

The above analytic theory of antiresonances was developed for
an idealized molecular wire model with semi-infinite single
channel leads. We now compare these analytic results with
numerical calculations for a more realistic molecular wire
model. The system we consider consists of (100) Au leads bonded
to a molecule as shown in the inset to Fig. \ref{fig1}.  It is
representative of a class of current experimental devices which
use a mechanically controlled break junction to form a pair of
nanoscale metallic contacts which are then bridged by a single
molecule, the molecular wire.\cite{Reed97} The molecular wire
we consider consists of two ``chain'' segments and an
``active'' segment.  The purpose of the chains is to reduce the
many propagating electron modes in the metallic contacts down
to a single mode which propagates along the chains.  Thus the
(finite) chains supplant the 1D ideal leads of our analytic
model. We model this molecular wire and its bonding to the 3D
metallic contacts by using extended H\"{u}ckel to calculate the
hopping elements and overlaps between the non-orthogonal atomic
orbitals that make up this system. It is the interaction
between the non-orthogonal orbitals on the chains and the
active segment that generates the antiresonances. Each chain
consists of 7 C-H groups and is terminated with a sulphur atom
which bonds to a gold lead.  For the energies of interest (near
the Fermi level of gold) these chains only support a single
$\pi$ mode. We chose an arbitrary active molecular segment with
two $\pi$-like MO's which only interact with the $\pi$ mode of
the chains.  The active segment is considered to be long enough
that there is no direct coupling between the chains, as in our
analytic theory. Fig. \ref{fig1} shows a  plot of the
contact-to-contact transmission calculated numerically for this
model. The arrows indicate the locations of the antiresonances
predicted by our analytic model (i.e., Eq. (\ref{eq:anti}))
using the same model parameters. The agreement is very good;
two antiresonances are found in each case at -10.2 eV and -10.4
eV, close to the Fermi energy of the gold leads.  The
transmission does not drop exactly to zero since in this
calculation second nearest neighbor interactions are also
included.  The experimental signature is a drop in the
differential conductance of the molecular wire. The agreement
between Eq. (\ref{eq:anti}) and Fig.  \ref{fig1} indicates
that our analytic result derived for the idealized molecular
wire model using the new approach to take account of
non-orthogonality has predictive power for more
complex systems.

In conclusion: Many of the quantum problems that arise in
physics and chemistry are formulated most naturally in a basis
of states that are not mutually orthogonal. In this Letter we
have shown that the exact solution of such problems is greatly
facilitated by embedding these non-orthogonal basis states in a
new Hilbert space in which they are by definition mutually
orthogonal but the matrix elements of the Hamiltonian are
energy-dependent. The power, simplicity and flexibility of this
novel approach was illustrated by applying it to analytic and
numerical calculations of electronic quantum transport in
molecular wires. A new mechanism for molecular wire conductance
antiresonances was identified which arises solely out of the
non-orthogonality of local orbitals on different sites of the
wire.

We would like to thank H. Trottier for rewarding
discussions. This work was supported by NSERC.

\end{multicols}


\begin{references}
\bibitem{var} If the basis $\lbrace |n\rangle \rbrace$ is
incomplete, Eq. (\ref{eq:Schrodbasis}) may be understood
variationally. See, for example, Ref.\cite{Lowd50}.
\bibitem{hilbert}Let $V$ be the vector space
spanned by the basis vectors $\lbrace |n\rangle
\rbrace$, in which $\langle m |n\rangle=\delta_{m,n}\,$.
The new Hilbert space $A'$ is the completion of $V$
with respect to the norm topology.
\end{references}
\end{document}